\documentclass[aps,prl,twocolumn,superscriptaddress,groupedaddress]{revtex4}  
\usepackage{graphicx}  
\usepackage{dcolumn}   
\usepackage{bm}        
\usepackage{amssymb}   

\begin{document}



\title{Hyper-Structured Illumination}

\author{Evgenii Narimanov}

\affiliation{School of Electrical and Computer Engineering, and Birck Nanotechnology Center, \\ Purdue University, West Lafayette IN, 47906}


\begin{abstract}
We present a new approach to super-resolution optical imaging, based on
structured illumination in hyperbolic media. The proposed system allows for
planar geometry, has unlimited field of view, and  is robust  with respect to optical noise and material losses.
\end{abstract}

\maketitle

\section{Introduction}
Fueled primarily by the promise \cite{Pendry2000} of optical imaging beyond the diffraction limit,  metamaterials, \cite{ref:metamaterials} artificial composites with the unit cell size well below the optical wavelength, now for more than a decade remain one of the primary foci of modern research. However, while  metamaterial devices, from the originally promised superlens \cite{Pendry2000} based on the negative index media, to its younger challenger the hyperlens \cite{OE2006,Engheta2006} that uses hyperbolic metamaterials,  have been demonstrated in experiment \cite{ZhangHyperlens, SmolyaninovHyperlens}  in broad range of frequencies and even extended to other wave phenomena such as e.g. ultrasound imaging, \cite{UltrasoundHyperlens} they have yet to find their way to broad acceptance in real-world applications. 

The inherent curvature of the metamaterial device that is necessary for optical magnification (either in the general formfactor of the device \cite{OE2006,Engheta2006,MagnifyingSuperlens} or in its unit cell topology \cite{PlanarHyperlens}), leading to the requirement on precise positioning of the object at subwavelength scale, remains a major practical obstacle to a wider acceptance of metamaterials-based imaging systems. More important, however, is the result of  substantial loss in metamaterial structures, that is generally a direct and inevitable consequence of using metallic (or other negative permittivity) components that remain necessary for either the negative-index metamaterials or hyperbolic media. Faced with  an order of magnitude intensity loss \cite{ZhangHyperlens} as the cost of producing a magnified  optical image of a subwavelength pattern, one must either accept a prohibitive cost in the signal-to-noise ratio, or substantially increase the intensity of the field that illuminates the target. The latter option however does not fit well with the constraints of biomedical imaging, where high optical fields may severely damage biological tissue.  

While directly related to the fundamental restrictions on optical metamaterials \cite{Stockman}  via the Kramers-Kronig relations, this problem however allows a natural and straightforward solution: the metamaterial component of a super-resolution imaging setup must be located ``in front'', rather that ``after'', the object -- so that it's probed by the already attenuated field, thus avoiding damage. This is precisely the case in the structured illumination \cite{Gustaffson} approach to optical imaging, where  a grid pattern, usually generated through the interference of incident light, is superimposed on the specimen.\cite{Gustaffson,si} The imaging resolution  is then determined by the periodicity of the illuminating field, and for illumination from free space this limit 
 reduces to one quarter of the wavelength, $\Delta = \lambda_0/4$.
 However, as we demonstrate in the present paper, hyperbolic metamaterials that are free from the conventional diffraction limit and can support optical patterns with the periodicity that is orders of magnitude below the free-space wavelength $\lambda_0$, can dramatically improve the imaging resolution in structured illumination. The resulting   {\it hyper-structured illumination} approach to super-resolution imaging is the main objective of this work.

\section{Structured Illumination in Hyperbolic Media}

In a conventional transparent medium with the refractive index $n$, the frequency $\omega$ generally limits the magnitude of the wavenumber $k$ to the value of $n \omega/c$. However, the situation is  dramatically different in the world of hyperbolic metamaterials, \cite{OE2006} where
the opposite signs of the dielectric permittivity components in two orthogonal directions ($\epsilon_\tau \epsilon_n < 0$ -- see the inset in Fig. 1(a))  lead  to a hyperbolic dispersion of $p$-polarized propagating waves
\begin{equation}
\frac{k_\tau^2}{\epsilon_n} + \frac{k_n^2}{\epsilon_\tau}  =  \frac{\omega^2}{c^2},
\label{eq:hyperbola}
\end{equation}
with the wave numbers unlimited by the frequency. For $k \gg \omega/c$, the hyperbolic iso-frequency surface of (\ref{eq:hyperbola}) asymptotically approaches a cone, with
$
k_n \simeq \sqrt{\frac{\epsilon_\tau}{\epsilon_n}} k_\tau.
$

As the direction of the group velocity ${\bf v}_g$ corresponding to a given wavevector ${\bf{k}}$ coincides with the normal to the iso-frequency surface, at large wavenumbers in a uniaxial hyperbolic medium  the angle between the group velocity and the material 
symmetry axis $\hat{{\bf z}} \equiv \hat{\bf{n}}$ (see the inset to Fig. 1(a)) approaches the value of $\theta_c = \arctan\sqrt{{\epsilon_\tau}/{\epsilon_n}}$. As a result, for  a point source placed at the edge of such medium, the emission intensity diagram forms a conical pattern (see Fig. 1 (b)-(d), calculated for a practical realization of hyperbolic media using silver-glass layered metamaterial), with the ``thickness'' of the expanding conical ``sheet'' of electromagnetic radiation well below the free-space wavelength. This phenomenon is now well-established, and has recently been successfully used for subwavelength optical lithography. \cite{hyper-lithorgraphy}   

As strong material dispersion is inherent to hyperbolic materials, a change in the frequency $\omega$ will result in a substantial variation of the emission cone angle $\theta_c$, as illustrated in Fig. 1(a) for three different wavelengths. While generally considered detrimental to metamaterial applications (since this would immediately push the system away from whatever resonant condition is was designed for), it is precisely this property that is essential to the concept of the hyper-structured illumination. While a single source (or an opening) at the bottom of the hyperbolic ``substrate'' only illuminates a small part of the object plane at its ``top'', a change in the electromagnetic frequency would allow to sweep the entire object plane (see Fig. 1 (b) - (d)). As a result, a complete subwavelength image can be obtained in a  single hyper-spectral measurement. 

This behavior is further illustrated in Fig. 2, which shows the object formed by four silicon nanowires on top of the silver-glass metamaterial ($10$ silver and $11$ glass layers, each $5$ nm  thick  -- see Fig. 2 (a)), illuminated at the wavelengths of $425$ nm (panel (b)) and $575$ nm (panel (c)). Note the subwavelength localization of the electromagnentic field near the object, and a  dramatic difference in the intensity distribution for different wavelengths that allows to clearly ``resolve'' distinct elements of the target despite their close spacing.

A finite size of the unit cell of the metamaterfial, $a$, introduces an upper limit on the (Bloch) wavenumber $k_n \equiv k_z$, equal to $k_n^{\rm max} = \pi/a$, as well as quantitative corrections to the dispersion $\omega\left({\bf k}\right)$ and to the geometry of the iso-frequency surface,  leading to its deviations from the hyperbolic shape. However,  the overall topology of the iso-frequency surface, together with  its dependence on the wavelength, are generally preserved -- see Fig. 1 (a).  As a result, one will still observe the general profile of the conical illumination pattern,  as well as its evolution with  the wavelength that are consistent with the predictions of the effective medium theory, as clearly seen in Fig. 1 (b)-(d).  However, as the size of the unit cell limits the maximum wavenumber that can be supported by the propagating waves in the hyperbolic metamaterial, it defines the ultimate resolution limit $\Delta$ of the hyper-structured illumination:
\begin{equation}
\Delta \simeq a
\label{eq:hsi_limit}
\end{equation}
With the scale of the unit cell in high-quality hyperbolic metamaterials that have already been demonstrated, on the order of a few nanometers, \cite{Liu} Eqn. (\ref{eq:hsi_limit}) allows for optical imaging of deeply subwavelength objects, to the point of resolving large individual molecules.

\begin{figure}[htbp]
\centering
\fbox{\includegraphics[width=3 in]{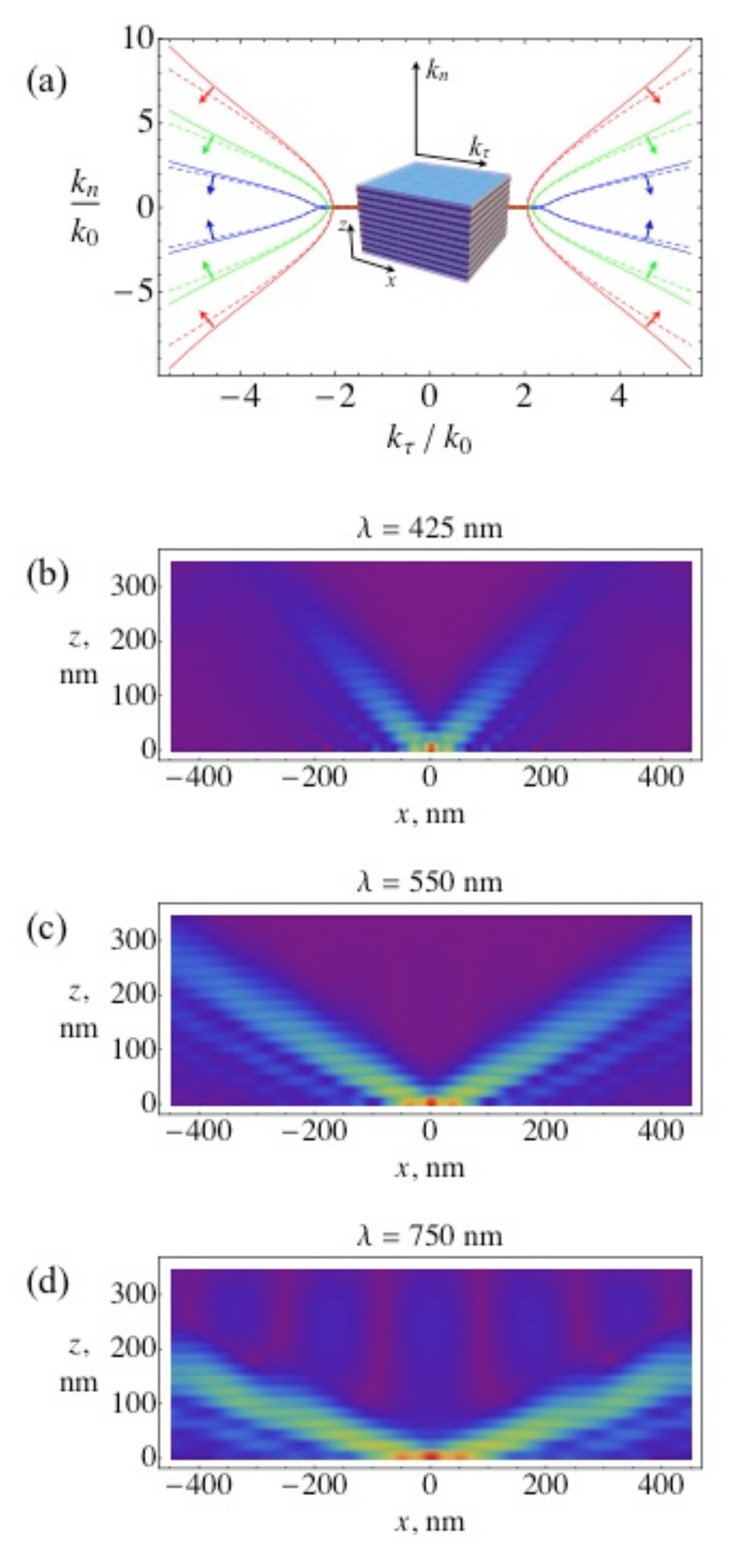}}
\caption{Panel (a): iso-frequency curves in a silver-glass hyperbolic metamaterial, for three different free-space wavelengths $\lambda_0$ : 425 nm (blue curves), 550 nm (green) and 750 nm (red). Solid curves are obtained from the exact calculation taking into account the finite width and the actual material loss of each layer ($a = 5 \ {\rm nm}$), while the dashed lines correspond to the effective medium approximation (see Eqn. (\ref{eq:hyperbola})). The arrows normal to the iso-frequency curves, indicate the directions of the group velocity. The inset shows the schematics of the silver-glass hyperbolic metamaterial, and introduces the coordinate system.
 Panels (b)-(d) : the magnetic field intensity for light of different wavelengths ($\lambda_0 = 425 \ {\rm nm}$ (b), $ 550 \ {\rm nm}$ (c) and $ 750 \ {\rm nm}$ (d)), incident from $z < 0$ half-space onto the hyperbolic metamaterial formed on top of a thick silver film with a narrow slit centered at $x = 0$. The slit is  parallel to the $y$-axis and is $5 \ {\rm nm}$ wide in the $x$-direction.  Note the conical emission pattern typical for any point source in  a hyperbolic medium, and strong sensitivity of its directionality to the wavelength.
 }
\label{fig:1}
\end{figure}

\begin{figure}[htbp]
\centering
\fbox{\includegraphics[width=3.25 in]{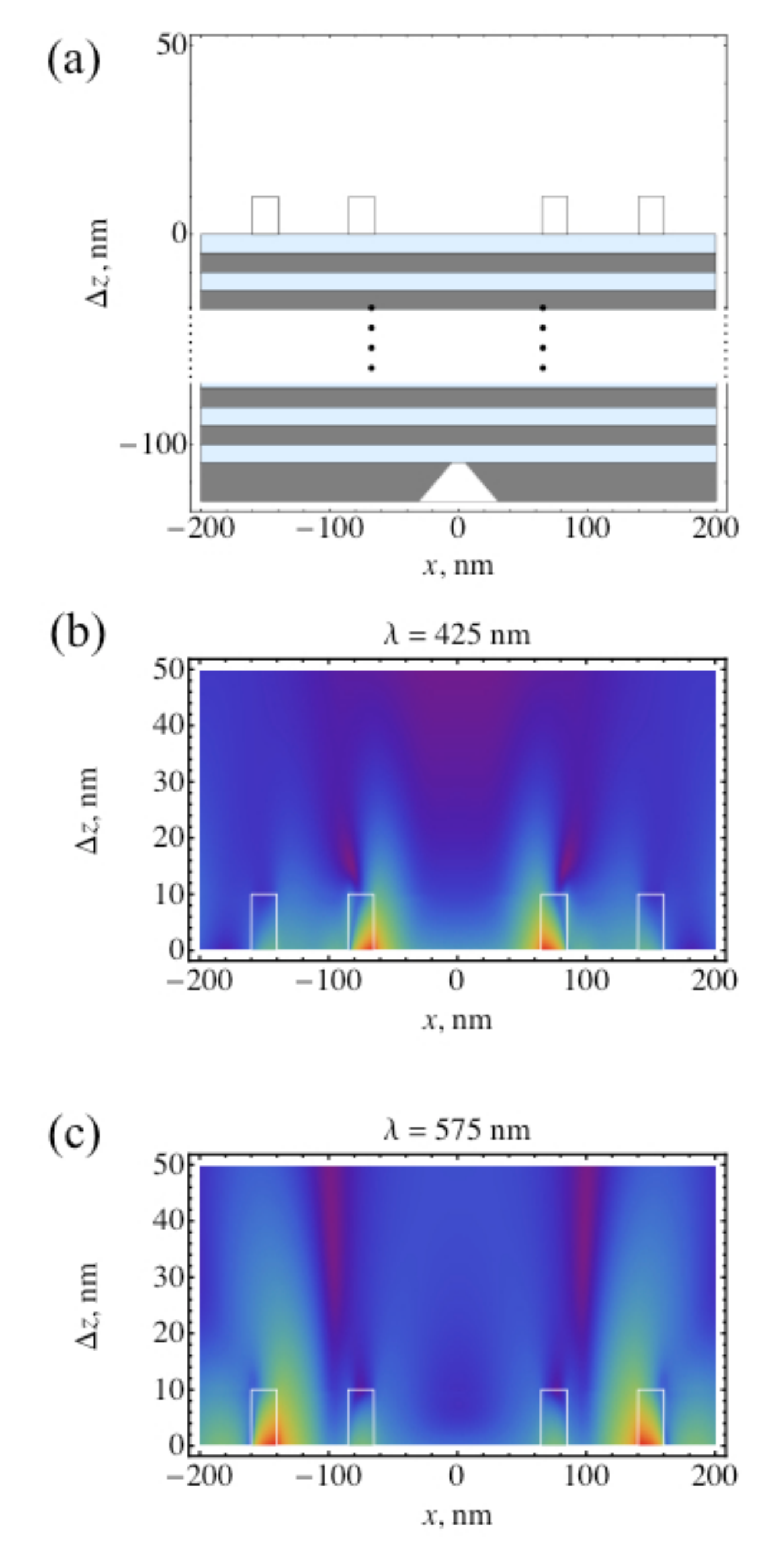}}
\caption{Hyper-structured illumination, with silver-glass hyperbolic metamaterial substrate.
 Panel  (a) shows the geometry of the imaging system, with dark and light-gray regions corresponding to silver and glass layers (each $5$ nm thick) respectively. Depending on the illumination wavelength, light incident from the slit at the bottom of the metamaterial, illuminates different areas of the target -- see panels (b) and (c), with a sub-diffraction resolution. The imaging target consists of four long silicon nanowires with rectangular cross-section of $20 \ {\rm nm} \times 10$ nm, aligned along the $y$-direction. The false-color representation shows the magnitude of the magnetic field, calculated taking full account of the material absorption in the system.}
\label{fig:2}
\end{figure}

\section{Depth of field in hyper-structured illumination and the Motti projection}

While a hyperbolic metamaterial supports propagating electromagnetic waves with the characteristic length scale up to its unit cell size, these waves will evanescently decay outsize this medium. As a result, the imaging resolution of the proposed hyper-structured illumination setup will rapidly deteriorate with  $\Delta z$ (see Fig. 2(a)), approaching the diffraction-limited value of $\lambda_0 / 4$ for $\Delta z \gg \lambda_0$. For a 3D object structured at a deeply subwavelength scale, the proposed imaging system based on hyper-structures illumination, will resolve the Motti projection \cite{MorseFeshchbach}  $\Delta_M\epsilon\left(x,y\right)$ of its tree-dimensional permittivity distribution $\epsilon\left({\bf r}\right)$:
\begin{eqnarray}
\Delta_M\epsilon\left(x,y\right) & = & \int \frac{d^3{\bf r'} \cdot \left( z' / 2 \pi\right) \cdot \left(\epsilon\left({\bf r'}\right) - 1\right)  }{\left(\left(x - x'\right)^2 + \left(y - y'\right)^2 + z'^2\right)^{3/2}}.
\label{eq:Motti} 
\end{eqnarray}
Note that the integral transformation in Eqn. (\ref{eq:Motti}) is independent of the illumination wavelength, and does not introduce an additional length scale: the effective averaging in (\ref{eq:Motti}) is defined by the distance $z'$ to the object plane. As a result, Eqn. (\ref{eq:Motti}) reduces the resulting resolution $\Delta$ to the corresponding depth of fields $\Delta z$.
This is illustrated in Fig. 3, which shows the Motti projection (panel (b)) for several solid spheres near the object plane $\Delta z = 0$.

\begin{figure}[htbp]
\centering
 \fbox{\includegraphics[width=2.5 in]{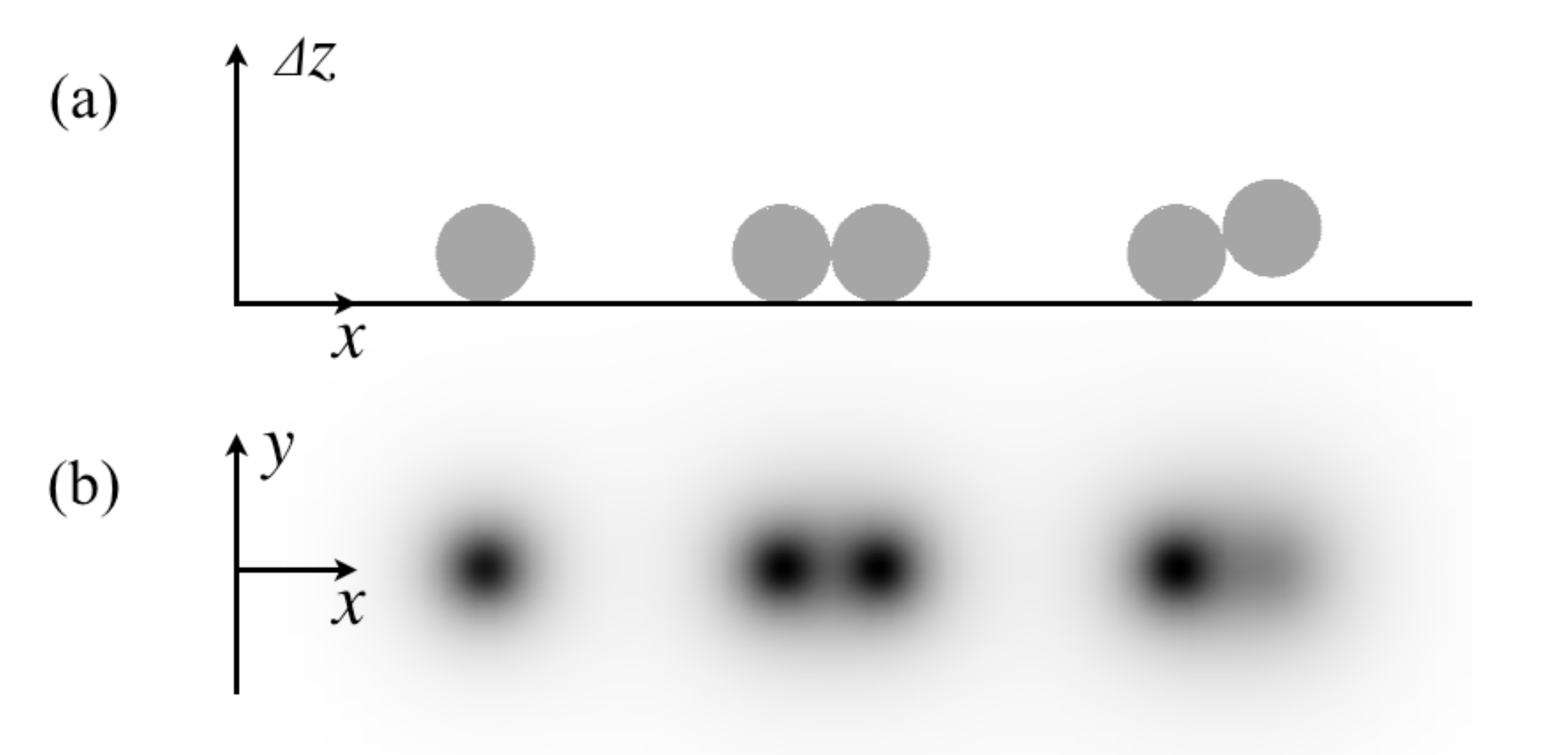}}
\caption{A group of five identical solid spheres positioned near the object plane (panel (a)) and their Motti projection (panel (b)), shown in gray scale. Note the reduced
contribution to $\Delta_M\epsilon\left(x,y\right)$ at longer separation from the object plane $\Delta z = 0$.} 
\label{fig:3}
\end{figure}

\section{ Theoretical Framework}

When illuminated from the slits at the ``bottom'' of the hyperbolic substrate (see Fig. \ref{fig:2}(a)),  
an object with the dielectric permittivity $\epsilon\left({\bf r}\right)$ placed near the image plane $\Delta z = 0$, will
induce scattered light with the far-field amplitudes for the $s$ and $p$ polarizations
\begin{eqnarray}
E_s\left({\bf k}; \omega\right) & =& \frac{4 \pi^2 \omega^2}{c^2 k_n k_\tau} \left\{ \left( {\bf p}_{\bf k}  \cdot  \left[\hat{{\bf n }} \times {\bf k}\right]  
\right) \right. \nonumber \\ 
& + & \left. r_s\left({\bf k'}\right)  \left( {\bf p}_{\bf k'}  \cdot  \left[\hat{{\bf n }} \times {\bf k'}\right]  
\right)   \right\}, \label{eq:mtr_s}\\
E_p\left({\bf k}; \omega\right) & =& \frac{4 \pi^2 \omega}{c k_n k_\tau} \left\{ \left( \left[{\bf p}_{\bf k} \times {\bf k}\right] \cdot  \left[\hat{{\bf n }} \times {\bf k}\right]  
\right) \right. \nonumber \\ 
& - & \left. r_p\left({\bf k'}\right) \left( \left[{\bf p}_{\bf k'} \times {\bf k'}\right] \cdot  \left[\hat{{\bf n }}\times {\bf k'}\right] \right)  \right\},
\label{eq:mtr_p}
\end{eqnarray} 
where $k_\tau$ and $k_n$ correspond to the tangential and normal to the surface of the hyperbolic substrate components of the far-field wavevector 
${\bf k} \equiv \left({\bf k}_\tau, {\bf k_n}\right)$,  $\hat{\bf n}$ is the  normal to the surface unitary vector,  $r_s$ and $r_p$ are the reflection coefficients at the air-substrate interface for the $s$- and $p$-polarizations,   ${\bf k'} \equiv  \left({\bf k}_\tau, - {\bf k_n}\right)$, and ${\bf p}_{\bf k}$ is the spatial Fourier transform of the polarization of the object, ${\bf p}\left({\bf r}\right) \equiv (1/4\pi) \left(\epsilon\left({\bf r}\right) - 1\right){\bf E}_i\left({\bf r},\omega\right)$, when it is illuminated by the incident field ${\bf E}_i\left({\bf r},\omega\right)$. For an object placed close to the image plane and with the vertical dimension well below the free-space wavelength ($\Delta z \ll \lambda_0$), the polarization of the object
${\bf p}_{\bf k}$ can be related to the spatial Fourier transform $\Delta_M\epsilon\left({\bf q}\right)$ of its Motti projection (\ref{eq:Motti}) :
\begin{eqnarray}
{\bf p}_{\bf k} & = & \frac{1}{8 \pi^2} \int d^2{\bf q} \ {\bf E}_i\left({\bf k + q};\omega\right) \ \Delta_M\epsilon\left({\bf q}\right).
\end{eqnarray}
Given the hyper-spectral measurement of the far field  ${\bf E}\left({\bf k}, \omega\right)$, the system of coupled linear equations (\ref{eq:mtr_s}),(\ref{eq:mtr_p}) can be used to  calculate $\Delta_M \epsilon$.
In particular, when the hyper-structured illumination pattern  ${\bf E}_i\left(x,y\right)$ at the image plane $\Delta z = 0$ originates from a periodic 
array of the slits at the bottom of the substrate, separated by the distances of $d_x$ and $d_y$ in the $x$ and $y$ directions,
\begin{eqnarray}
{\bf E}_i\left(x,y; \omega\right) =   \sum_{m_x, m_y = - \infty}^{\infty} 
{\bf E}_0\left(x - m_x d_x, y - m_y d_y; \omega \right),  \ 
\end{eqnarray}
then each spatial Fourier component of the object  profile $\Delta \epsilon\left({\bf q}\right)$ is only coupled to a discrete set of  other spatial components, that are different by an integer multiple of the reciprocal lattice vectors $2\pi/d_x \ \hat{{\bf x}}$ and 
$2\pi/d_y \  \hat{{\bf y}}$:
\begin{eqnarray}
\sum_{n_x, n_y = - \infty}^{\infty} G^{(s,p)}_{n_x,n_y}\left( {\bf k}; \omega \right) \ \delta\epsilon_{n_x,n_y} 
 = 
\frac{2 c k_n k_\tau} { \omega}  E_{(s,p)}\left( {\bf k}; \omega \right) \ 
\label{eq:InvMtr}
\end{eqnarray}
where 
\begin{eqnarray} 
\delta\epsilon_{n_x,n_y} \equiv \Delta_M\epsilon\left({\bf k}_\tau + \hat{\bf x} \ 2\pi  \ n_x / d_x +  \hat{\bf y} \ 2 \pi n_y / d_y   \right),
\end{eqnarray}
 and 
\begin{eqnarray}
G^{(s)}_{n_x,n_y}\left( {\bf k}; \omega \right) & = & \left(1 + r_s\right) \left( \left[\hat{{\bf n }} \times {\bf k}\right]   \cdot  { \bf  e}_{n_x,n_y}\left(\omega\right) \right), \nonumber \\
G^{(p)}_{n_x,n_y}\left( {\bf k}; \omega \right) & = & \frac{c}{\omega} \  \left[\hat{{\bf n }} \times {\bf k}\right]  \cdot    
 \left[ { \bf   e}_{n_x,n_y}\left(\omega\right) \times  \left({\bf k} - r_p \ {\bf k'}  \right)
\right], \nonumber
\end{eqnarray}
with
\begin{eqnarray}
 { \bf   e}_{n_x,n_y}\left(\omega\right) \equiv \frac{1}{d_x d_y} \int d^2{\bf r} \ {\bf E}_0\left({\bf r};\omega\right)
 \ e^{ \ 2\pi  i \left( x n_x / d_x +   y n_y / d_y   \right) } .
 \nonumber
\end{eqnarray} 
Given the desired resolution $\Delta$, this implies a summation by $n_{\rm max} \sim d/\Delta$ terms in either direction. Even with $\Delta \sim 10$ nm and diffraction-limited slit-to-slit spacing $d \sim \lambda_0/2$, we find  $n_{\rm max} \sim 10$ -- thus reducing the numerical complexity of solving Eqn. (\ref{eq:InvMtr}) to that of inverting a $100 \ {\rm x} \ 100$ matrix. On a modern CPU, such computational task can be completed within a 0.1 ms time frame. Thus, despite the need for data post-processing that is inherent to structured illumination methods, the proposed approach can be 
successfully used for real-time imaging of the dynamical processes at sub-ms time scales.

\section{Hyper-structured Illumination in a noisy environment}

The  linear inversion procedure of the previous  section is robust to the presence of the noise and material absorption -- which is illustrated in Fig. 4, where panels (a) and (b) present the gray-scale images of the original 
object and its image obtained in a noisy environment. For a quantitative comparison, panel (c) shows the dimensions of the original object (left scale) and the  recovered Motti projection (right scale). Note that the $30$ nm separation, which is less than  $1/10$ of the shortest illumination wavelength, is clearly resolved -- despite a substantial amount of noise in the system.

\begin{figure}[htbp]
\centering
\fbox{\includegraphics[width=2.75 in ]{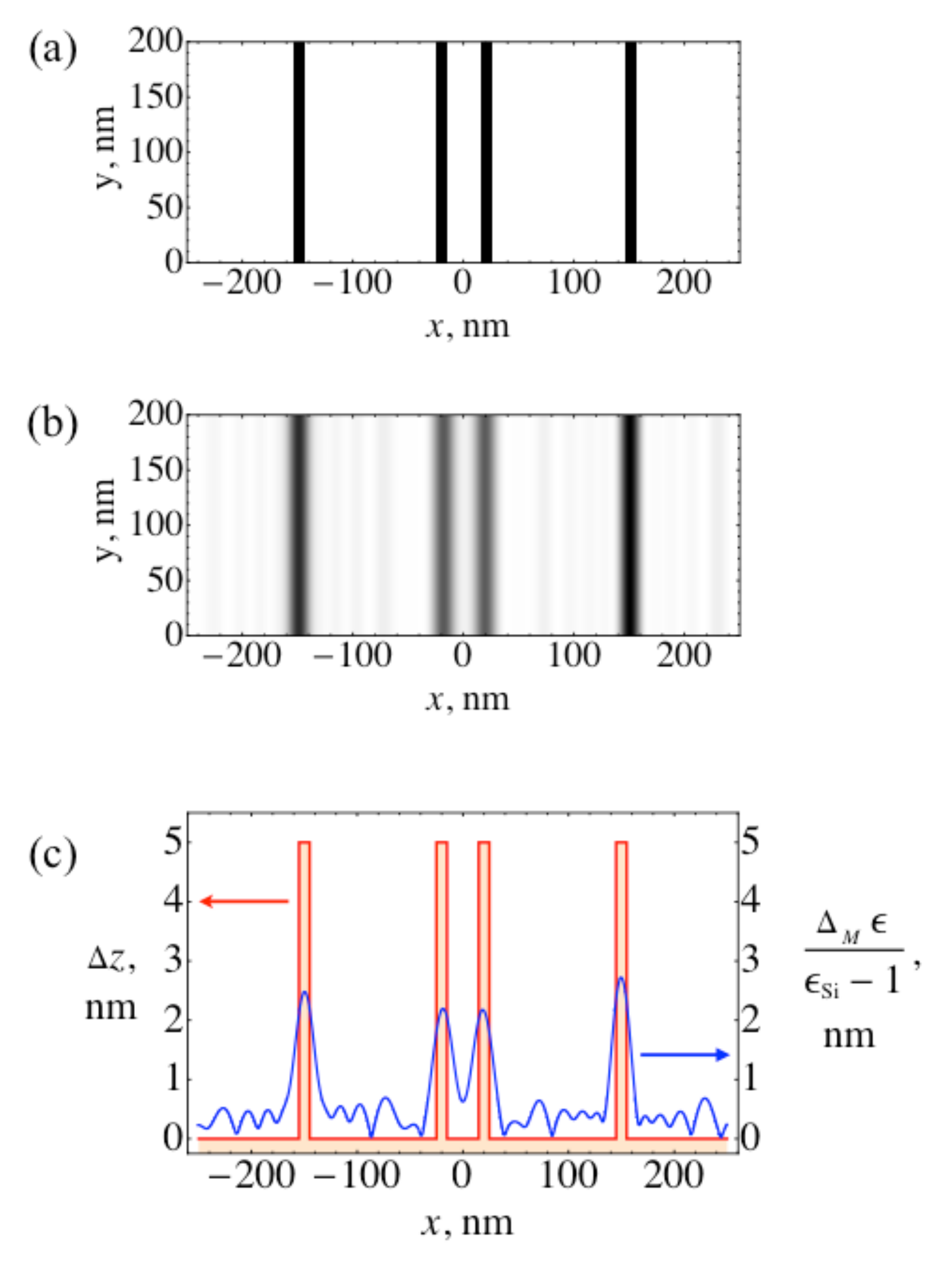}}
\caption{Super-resolution imaging with hyper-structured illumination. Panel (a) : the ``top view'' of the object  (four identical silicon
nanowires with 10 x 5 nm cross-section) on the silver-glass hyperbolic substrate (see Fig. 2(a)). Panel (b) : the reconstructed image, 
in gray-scale. The calculation includes the effects of the actual material loss in the silver-glass hyperbolic substrate. Panel (c): the object profile (red) and the  (normalized to the silicon permittivity 
$\epsilon_{\rm Si}$)  reconstructed Motti projection $\Delta_M\epsilon /( \epsilon_{\rm Si} - 1)$ (blue).}
\label{fig:4}
\end{figure}

\section{Discusson and Conclusions}

The hyper-structures illumination approach introduced in the present paper, combines deeply subwavelength resolution
with the relative simplicity of {\it planar} geometry that would facilitate surface functionalization for cell targeting and imaging, essential for biomedical applications. Compared to the existing super-resolving extensions of structured illumination, such as the 
plasmonic structured illumination \cite{pSI}, the proposed approach is not limited to particular resonant wavelength (such as e.g.
the surface plasmon wavelength), and is not sensitive to the material absorption.

In terms of the actual fabrication of the required metamaterial substrate, it's well within the limits of already demonstrated 
capabilities, \cite{Liu} and experimental demonstration of the proposed imaging method will be straightforward.

$ { \ } $

This work was partially supported by NSF  (grant 1120923) and  Gordon and Betty Moore Foundation.





\end{document}